\documentclass[superscriptaddress,aps,pra,twocolumn,preprintnumbers,amsmath,amssymb,floatfix,groupedaddress]{revtex4}

\usepackage[dvips]{graphicx}% Include figure files
\usepackage{dcolumn}% Align table columns on decimal point
\usepackage{bm}% bold math
\usepackage{amssymb}
\usepackage{xr}

\usepackage{color}

\newcommand{\cst}[1]{\ensuremath{\mathrm{#1}}}%typographie des constantes ?
\newcommand{\e}{\cst e}

\begin{document}

\title{Generating non-Gaussian states using collisions between Rydberg 
polaritons }

\author{ Jovica Stanojevic$^1$, Valentina Parigi$^1$, Erwan Bimbard$^1$,  Alexei Ourjoumtsev$^1$, Pierre Pillet$^2$ and Philippe Grangier$^1$}
\affiliation{
$^1$Laboratoire Charles Fabry, Institut d'Optique, CNRS, Univ Paris-Sud,  2 Avenue Fresnel, 91127 Palaiseau, France.  \\
$^2$Laboratoire Aim\'e Cotton,  B\^atiment 505, Univ Paris-Sud, 91405 Orsay cedex, France}

\date{\today}

%%%%%%%%%%%%%%%%%%%%%%%%%%%%%%%%%%%%%%%

\begin{abstract}
We investigate theoretically the  deterministic generation of quantum states  
with negative Wigner functions, by using giant  
non-linearities due to  collisional  interactions between Rydberg polaritons. 
The state resulting from the polariton interactions may be transferred with  high fidelity  into a  photonic state, 
which can be analyzed using homodyne detection followed by quantum
tomography. 
Besides generating highly non-classical states of the light, this method can also provide a very sensitive probe for  the 
physics of the collisions involving Rydberg states. 
\end{abstract}

%%%%%%%%%%%%%%%%%%%%%%%%%%%%%%%%%%%%%%%%
%\keywords{Suggested keywords}
%Use showkeys class option if keyword
%display desired

\maketitle

%\section{Introduction}

The generation and characterization of  highly non-classical states of the light have
accomplished considerable progress during recent years.
This includes for instance the production and analysis of
 one-  \cite{Hansen} and two-  \cite{Ourj2,Val}
photon Fock states, and of superpositions  of coherent states, often called 
``Schr\"odinger's cat" states \cite{cats}.  
In these experiments, the desired states are obtained  by using 
so-called ``measurement-induced non-linearities", where a measurement is performed
on one part of an entangled state. Then  the other part is  projected onto the desired state, 
conditional to obtaining the good result in the projecting measurement. 
This method works quite well, but it is intrinsically
non-deterministic~: the probability of success  is usually 
low, and  the desired state cannot be created ``on demand", when needed for
applications e.g. for quantum communications. 

\begin{figure}
\includegraphics[scale = 0.22]{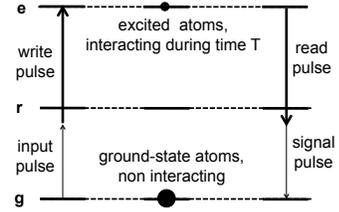}
\caption{
An ensemble of $N$ three-level atoms  
is initially prepared
in a coherent Rydberg polariton state, with only a few atoms excited. After a time
shorter than the decoherence time, the collisions have modified in a non-linear way 
the coefficients of the  polariton state, which is ``remapped" onto a photonic state,
and analyzed using an homodyne detection.  
\label{fig:scheme}}
\end{figure}

Here we would like to investigate another scheme, which can, at least in
principle, be made more deterministic, by using Rydberg interactions
in a cold atomic gas.  Rydberg states are highly excited atomic states, that interact very strongly at distances $R$ of order 
a few $\mu$m,  through either  dipole-dipole ($1/R^3$) or van der Waals ($1/R^6$) interactions \cite{Ry,rydbergs}. 
The idea is first to change a generic photonic state, 
for instance a weak coherent state, into a so-called ``polariton" state, where
a  long-lived excitation is
distributed over many atoms \cite{EIT}.  Here we will consider Rydberg polariton states, where many 
 atoms share a few delocalized Rydberg excitations \cite{Ry}.
The prepared polariton state will then evolve under the action of these Rydberg-Rydberg 
interactions, generating highly non-classical 
(typically non-Gaussian) states \cite{kuzmich}. When the desired state is obtained, the polariton state can be converted back into a  free-propagating photonic state,  using a control laser beam (see Fig.~\ref{fig:scheme}). 
The phase-matching condition between the input, write, read, and signal beams
leads to a collective enhancement effect ensuring that 
the  state of the light  is emitted in a well-defined spatial and temporal mode \cite{Bariani}. The  
conversion of the polariton state back to photons  
is then expected to  deterministically produce highly non classical states  \cite{Gorshkov07,Vuletic,Gorshkov11,us}. 
In addition, the analysis of the generated state will give 
information about the collisional mechanisms which take place between the Rydberg atoms. 

Let us emphasize that a particularly useful method  to fully characterize
highly non-classical  states is quantum tomography, which allows one   to
reconstruct the Wigner function $W(q,p)$ of the state in phase space, where
$\hat q$ and $\hat p$ are the quadrature operators of the quantized electric field, 
measured using homodyne detection \cite{Hansen,Ourj2,Val,cats}. 
Such a method provides a full characterization of the
measured quantum state, which is very intuitive because the ``non-classicality''
of the measured  state, and especially its purity, directly translate into
the negativity of the Wigner function \cite{Barbieri}. We will thus use 
such Wigner functions to characterize both  the photon and the polariton states. 

In this letter we will study  the first two steps of  such a  deterministic preparation of  non-Gaussian states~:

({\it i}) preparing the coherent polariton state using a weak laser pulse; we will  introduce 
the phase-matched symmetric 
Dicke states \cite{pillet} as a  convenient way to describe the states of the ground and excited atoms  
(this is essentially equivalent to the polariton picture \cite{EIT}).

({\it ii}) leaving the polariton state evolve for some time (shorter than its decoherence time). 
The state will then be modified in a non-linear way, depending on the nature of the Rydberg-Rydberg interactions.
We will obtain simple analytic expressions for this evolution, that are the main results of the present paper. 

Finally, we will  characterize the generated polariton states by computing    their Wigner functions, and  
we will  discuss  various experimental considerations, including the remapping of  the polariton into a photonic state.

%%%%%%%%%%%%%%%%%%%%%%%%%%%%%%%%%%%%%%%%%
%\section{Excitation of the atoms to collective Dicke states}
%%%%%%%%%%%%%%%%%%%%%%%%%%%%%%%%%%%%%%%%%

The calculation is performed by splitting  the evolution of the system in two steps : first, 
an excitation step using a short (typically $\sim$ 1 ns) and 
weak laser excitation pulse, creating a few Rydberg states; second, a free evolution of the 
generated state under the effect of Rydberg-Rydberg interactions.  We will show that we can 
consistently ignore the  interactions during the first  step, and then 
ignore the laser in the second step, since  it is turned off. 

In order to describe the excitation step for an ensemble of $N$ atoms from a ground state  $g$,   we consider 
a two--photon excitation, 
off-resonant from the intermediate level $r$,  and  resonant with the Rydberg state $e$ (see Fig. 1). 
It can thus be described using an 
effective two-level model \cite{rydbergs} with the  Hamiltonian $H=H_\ell + H_c$, where 
%%%%%%%%%%%%%%%%%%%%%%%%%%%%%%%%%%%%%%%%%%
\begin{eqnarray}\label{Hamiltonian}
H_\ell = \frac{\hbar \Omega(t) }{2} 
\sum\limits_{i = 1}^N  \left( {\hat \sigma _{eg}^i  +  
\hat \sigma_{ge}^i 
} \right), \; \;   
H_c = \sum\limits_{i = 1,j > i}^N {\hbar \kappa_{ij}} \hat \sigma
_{ee}^i \hat \sigma _{ee}^j .  \nonumber
\end{eqnarray}
%%%%%%%%%%%%%%%%%%%%%%%%%%%%%%%%%%%%%%%%%%
Here $\hat \sigma_{\alpha\beta}^i=\left|\alpha\right>\left<\beta\right|
\e^{\pm(1-\delta_{\alpha\beta}) i \vec{k}.\vec{R_i}}$, 
$\vec{R_i}$ is the position of atom $i$, and $\vec{k}$  the total  wave-vector of the exciting light, with 
an  effective (pulsed) Rabi frequency
$\Omega(t)$; $\hbar \kappa_{ij}$ is the  pair-wise interaction energy of the excited atoms, and 
$\delta_{\alpha\beta}$ the Kronecker symbol. 
 
For short times and very low excitation fractions, let us first neglect the interaction term $H_c$ and keep 
only the laser excitation $H_\ell$. 
The result of laser excitation is then a coherent polariton state, that is 
a superposition of  symmetric ``phase-matched" Dicke  states $\left|n \right>$,
where $n$ is the number of excited atoms (see Appendix 1). The amplitude $C_n$ corresponding to
the collective states $\left|n\right>$ directly follows from the
single-atom amplitudes
\begin{equation}\label{C_n-collect}
C_n =(-i)^n  \sqrt{B(N,n)}  \sin^n \frac{\omega }{2}
\cos^{N-n}\frac{\omega }{2},
\end{equation}
where $B(N,n)$ is the binomial coefficient, $N$ is the total number of
atoms and $\omega = \int \Omega(t) dt$  is the pulse area.  
In the limit of very small $n/N$, one has
\begin{equation}\label{C_n-ratio}
\lim_{n/N\rightarrow 0}   C_n/C_{n-1} =  \alpha/\sqrt{n},
\end{equation}
where
$\alpha= -i \sqrt{N} \tan(\omega/2)
\approx -i \sqrt{N} \omega/2$.
The amplitude $\alpha$ is related to the averaged number of
excited atoms $\langle N_{\rm exc}\rangle=N \omega^2/4=|\alpha|^2$.
Relations (\ref{C_n-ratio})  yields the well known
expression $C_n/C_0 = \alpha^n/\sqrt{n!}$
for the amplitudes of a coherent state, with $C_0$  determined by
normalization. For this calculation to be consistent, we need to fulfill  the 
condition  $\langle n | \exp{(-i H_c \tau/\hbar)} | n \rangle \sim 1$ at the end of the laser pulse of 
duration $\tau$, where $H_c$ 
is the collisional part of the Hamiltonian; this  will be checked in Appendix 2.

 %%%%%%%%%%%%%%%%%%%%%%%%%%%%%%%%%%%%%%%%%
%\section{Interactions between the  collective Dicke states}
%%%%%%%%%%%%%%%%%%%%%%%%%%%%%%%%%%%%%%%%%

After the laser pulse is off, we consider 
the evolution of the previous coherent Rydberg polariton
state under the only action of  the Rydberg interaction Hamiltonian $H_c$. 
A crucial  remark is that during 
the excitation and de-excitation phases (see Fig. 1), only the
the symmetric phase-matched Dicke states (see Appendix 1) will be
mapped coherently  between the polariton and photonic states \cite{EIT}. In addition, the  Hamiltonian $H_c$ 
preserves the number of excitations $n$, and therefore  it only mixes symmetric Dicke  states with non-symmetric ones,
which are uncoupled from the laser readout process. 

In order to characterize the phase-matched part of the state after an evolution time $T$, 
we need therefore to evaluate the matrix elements $\langle n |U|n\rangle$, where $U = \e^{-i H_c T/\hbar}$. 
For this purpose we use the following transformation
\begin{equation}\label{exponent-projector}
e^{-i \kappa_{pq}T\,\hat\sigma^{i_p}_{ee}\hat\sigma^{i_q}_{ee}} =1+
\hat\sigma^{i_p}_{ee} \hat\sigma^{i_q}_{ee}\left(e^{-i \kappa_{pq}T}-1\right).
\end{equation}
For a low number of excited  atoms, the probability to
have $p$ interacting atoms close to each other vanishes very quickly as $p$
increases. Therefore, the leading interaction order originates from
two-body interactions, the next-to-leading order originates from
three-body interactions and so forth. The transformation
(\ref{exponent-projector}) can facilitate this expansion because
the term $(e^{-i \kappa_{pq}T}-1)$ is zero if the two atoms do not interact, i.e.
if they are far from each other. One can then use 
these types of terms to select pairs of interacting atoms in various
interaction orders, and 
the expectation values $\langle n | U| n \rangle$ can be evaluated by
bookkeeping various combinations and contributions of $p$
excited atoms, with $p \leq n$. The expansion is finite (since $p\!\leq\!n$),
however the number of terms and their complexity rapidly
increase for $p\geq4$; we will therefore look first at low $n$, and then find an excellent ansatz for higher $n$.

Denoting
as $\eta({\bf r})$ the atom number density at point $\bf r$, and  $dN_i  = d^3r_i \;  \eta({\bf r_i})$, 
we define~: 
\begin{eqnarray}
&& I_2  \!=\! \frac{1}{N^2}  \iint dN_i \; dN_j \;  (e^{-i  \kappa_{ij} T}-1), 
\label{I21}  \\
&& J_3  \!=\! \frac{1}{N^3}  \iiint dN_i  \; dN_j  \; dN_s  \times \nonumber\\
 &&\hspace{7 mm} (e^{-i  \kappa_{ij} T}-1) (e^{-i \kappa_{is} T}-1) ,
\label{J33} \\
&& I_3 \!=\! \frac{1}{N^3} \iiint dN_i \; dN_j \; dN_s \times \nonumber \\
 &&\hspace{7 mm} (e^{-i  \kappa_{ij} T}-1) 
 (e^{-i \kappa_{is} T}-1) (e^{-i  \kappa_{js} T}-1) , \label{I33} %\\
\end{eqnarray}
and  one gets the successive terms
\begin{eqnarray}
\langle 0 |ÊU | 0 \rangle \!&=&\! \langle 1 |ÊU | 1 \rangle = 1, \nonumber \\
\langle 2 |ÊU | 2 \rangle\!&=&\! 1+ I_2, \nonumber \\
\langle 3 |ÊU | 3 \rangle\!&=&\! 1+3 I_2 +3  J_3\nonumber+\; I_3, \nonumber \\
\langle n |ÊU | n  \rangle \!&=&\! 1+B(n,2)I_2 +3 B(n,3) J_3 +\;B(n,3)I_3+\ldots  \nonumber
\label{expansion}
\end{eqnarray}
For each $\langle n |ÊU | n  \rangle$, the coefficients of  the quantities  $I_{2,3}$ and  $J_{3}$ correspond to the
numbers of different choices of excited atoms $(i,j)$ and $(i,j,s)$ that appear in 
 the expressions (\ref{I21})-(\ref{I33}).
Though this  approach is rigorous and can work in principle at any order, it
has the disadvantage that 
the integrals $I_n$ are more and more complicated to evaluate when $n$ increases.  
Whereas $I_2$ can easily be calculated analytically (see Appendix 3), 
this is more tedious for $I_3$, and $I_{n > 3}$ are only numerical. We therefore introduce now 
a much simpler approach, giving analytical results at any order, which works
surprisingly well when compared with numerical calculations. 

The idea  is to evaluate $U_{n>m}\,$, assuming  that  $U_m\!=\!\langle m|U|m\rangle$ is known.
For this purpose, we first note that the value of $U_n$ for a set $\{ n \} = \{i_1, \; i_2, \; ... \; i_n \}$ of $n$ excited atoms 
can be formally divided into a product of terms involving set $\{ m \} $ of $m < n$ atoms, 
in the following way (we note that the total number of excited atoms is always conserved) : 
\begin{equation}\label{prod-Hm}
 \left\langle \exp(-iK_c )\right\rangle_{\{n\} }\!\!=\!\!
\prod\limits_{i_1<...<i_m\atop i_m\leq i_n} \!\!\!\bigg\langle\!
\exp(-i \lambda_{nm} K_c )\!\bigg\rangle_{\{ m \} }
\end{equation}
where $ K_c= H_c T /\hbar$, and $\lambda_{nm}\!\!=\! B(n,2)/[ B(n,m)B(m,2) ]$ is due to the fact that each
$\kappa_{ij}$ appears in several subsets. 

We also note that the calculation of 
$U_m$ for a given set $\{ n \} $ of atoms will  involve an
average over all random
positions of atoms, that is essentially the continuum limit of the above expressions. 
Taking into account this averaging, the quantities 
$\langle \exp(-i\lambda_{nm} K_c ) \rangle_{\{ m \} }$ are the
same for all $m$-subsets.
As a consequence, we use as  a
crucial ansatz that the rhs of
Eq. (\ref{prod-Hm}) is just  a product of $B(n,m)$ identical factors 
$\langle \exp(-i\lambda_{nm} K_c )\rangle_{\{ m \} }$ :
\begin{eqnarray}
\langle n|U|n\rangle\!&=&\!\left\langle
\exp(-i K_c )\right\rangle_{\{n\} } \nonumber\\
&\approx & \,\langle \exp(-i K_c )\rangle_{\{m\}  }^{B(n,m) \lambda_{nm}}
\!=\!\langle m|U|m\rangle^{\frac{n(n-1)}{m(m-1)}} \; \; \; \; \; \label{H-scaling}.
\end{eqnarray}

For $m\!=\!2$, 
one has then simply:
\begin{equation}\label{U2-scaling}
\langle n |U | n  \rangle = (1+I_2(T))^{n(n-1)/2}
\end{equation}
where $I_2(T)$
can be calculated analytically
by integrating on the positions of two atoms within a sphere of  constant
atomic density, the result is
given in Appendix 3.

Figure \ref{nn} shows the plots of $ \langle n|U|n \rangle $, $n$ = 2, 3 and 4 for van der Waals interactions, as
a function of the scaled interaction time $t$. 
For each $n$ three different curves are almost perfectly overlapping~: 
the green dotted curves are the numerical results, the red full ones show the scaling from $n=2$
using the  expression of $I_2(t)$  
given in Appendix 3,  and  the blue dashed ones show the  scaling from $n=3$.
In the numerical calculation, groups of four atoms are generated and $\langle n|U|n \rangle $ is 
essentially  the averaged $U$ over random n-groups of atoms. 
The calculations are done for a sphere with a uniform density, but can be generalized for arbitrary density profiles. 
\begin{figure}
\includegraphics[scale =0.325]{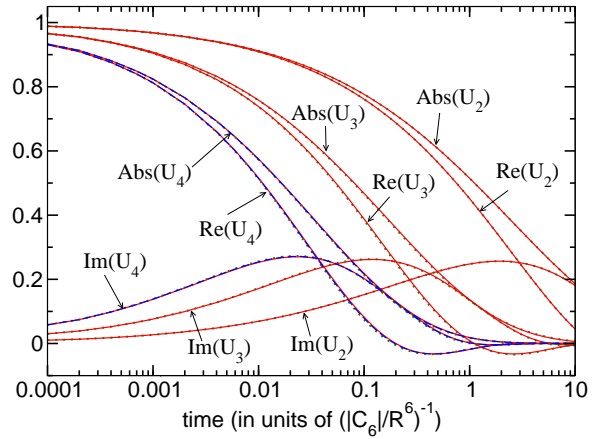}
\caption{Real and imaginary parts, and modulus of $U_m\!=
\!\langle m|U| m\rangle$ 
 as a function of time in units of $T_R=R^6/|C_6|$, for attractive van der
Waals interactions. For  repulsive interactions one has only  to change the sign 
 of ${\rm Im}(U_n)$. All red curves are obtained by scaling
from the corresponding $U_2$ using Eq. (\ref{U2-scaling}). All blue dashed 
lines are obtained using the scaling formula (\ref{H-scaling}) with the exact
numerical result for $U_3$, and 
all green dotted lines are the results of the full numerical calculation. 
Similar results (not shown here)  can be obtained for dipole-dipole interactions 
rather than Van der Waals  interactions. 
 \label{nn}}
\end{figure}

Summarizing, we have thus obtained a series of simple approximate expressions of
$\langle n | U | n \rangle$, valid for any $n$. 
This surprisingly simple derivation can be understood as an approximate but 
efficient way to resum the terms appearing  in the more
rigorous expansion quoted before.

\begin{figure}
\centerline{
\includegraphics[scale = 0.375]{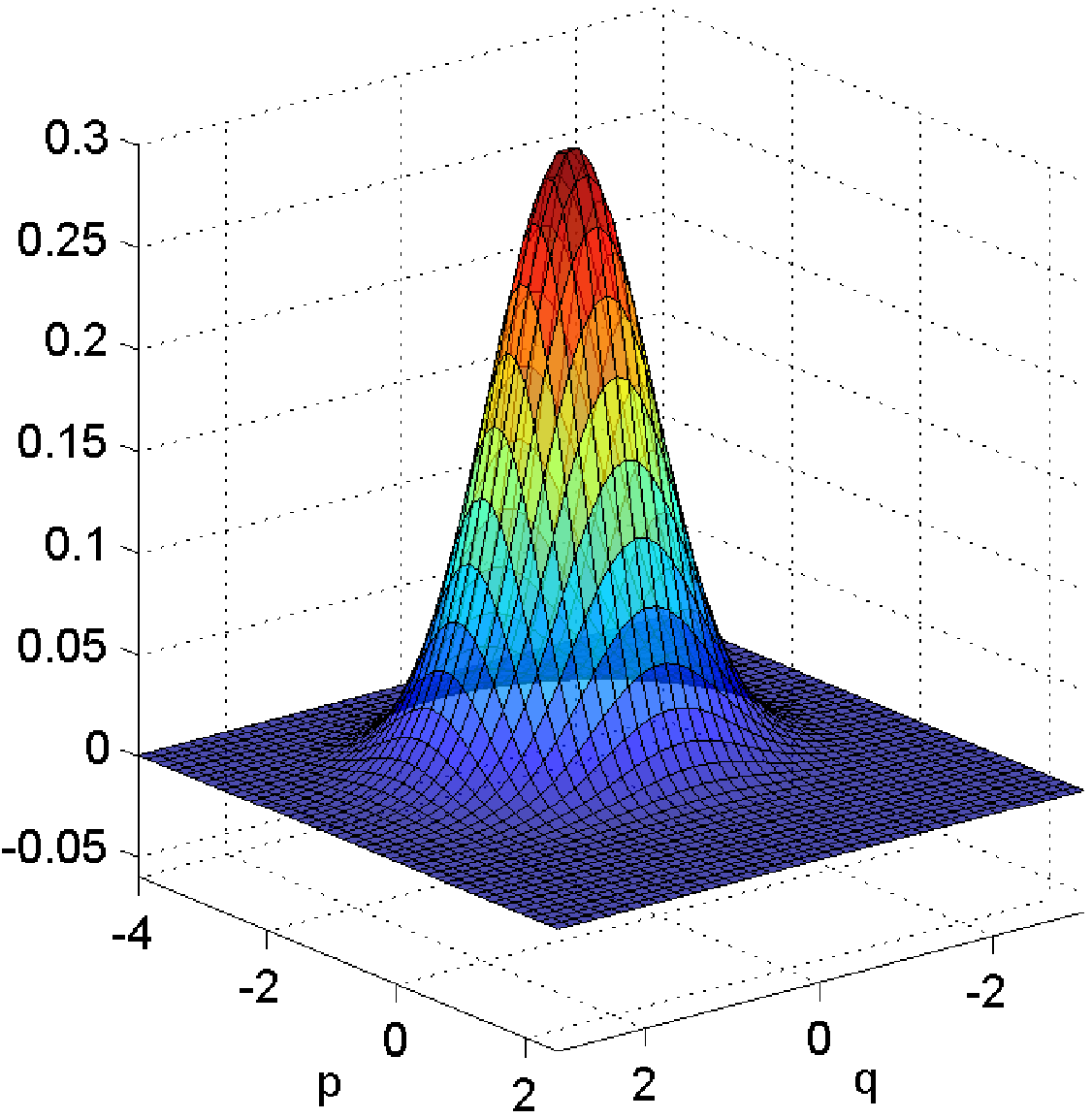}
\includegraphics[scale = 0.375]{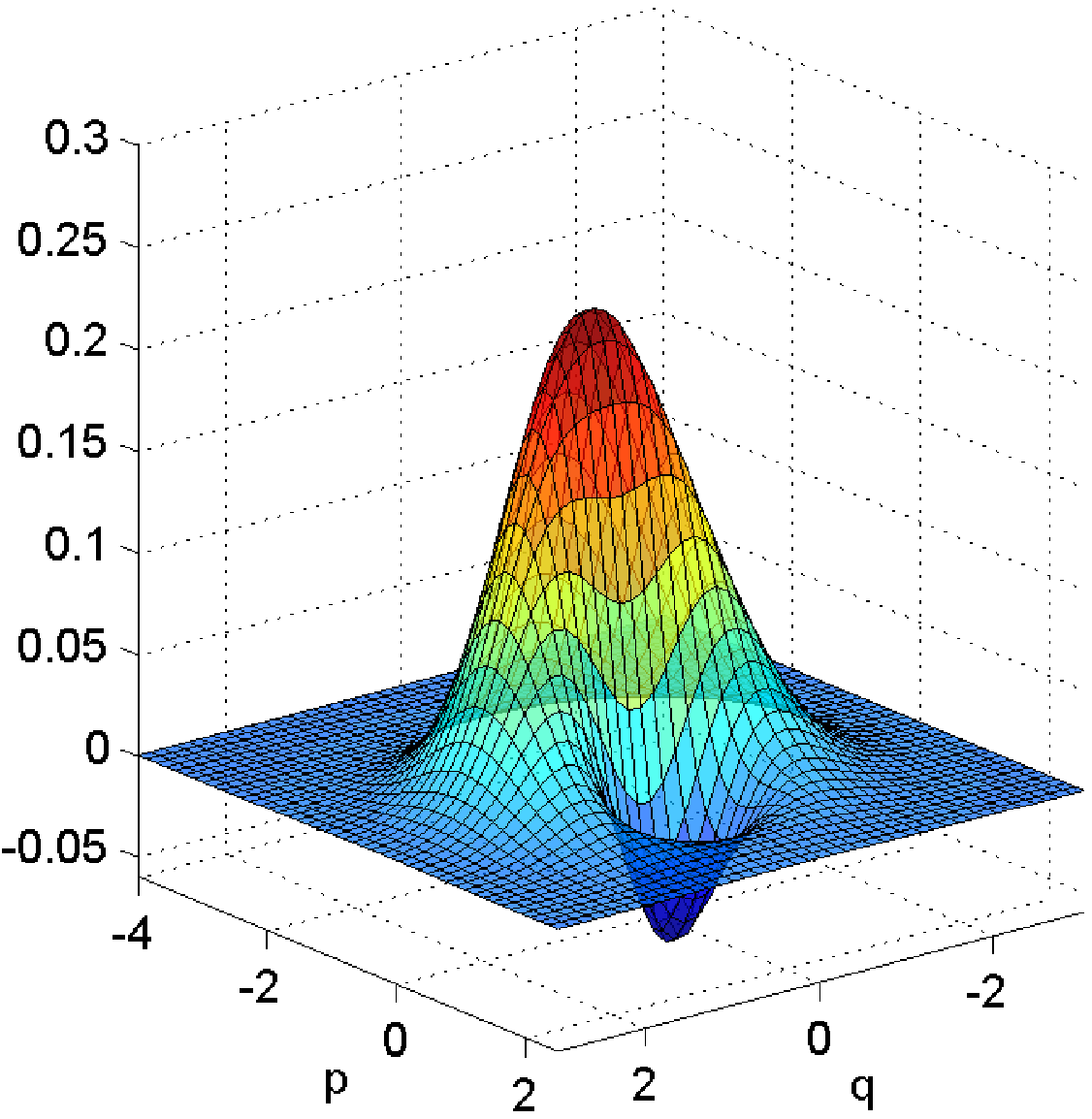}}
\caption{ 
Evolution of the Wigner function of an initial coherent state $ | \alpha \rangle$ with
$|\alpha| = 1$, between  the scaled times 0.004 (left side) and 4 (right side), that is  close
to the truncated state  $ | 0 \rangle  + \alpha | 1 \rangle$. 
The corresponding physical times are 0.72~ns (left) and 720~ns (right), 
with Van der Waals interactions  only. 
\label{wigner}}
\end{figure}

In order to analyze the generated polariton states, it is convenient to use
Wigner functions, that  show the evolution of the initial Gaussian  into
non-Gaussian states. Simple analytical calculations give easily  $W(q,p)$ in a
suitably truncated Fock state basis, as a function of the coefficients $\langle
n | \hat \rho | m \rangle $ of the density matrix, obtained from the evolution
of the initial Dicke state using previous formulas. 
For long interaction times, the result is rather simple : since the  coefficients for $n=0$ and 1 are 
unaffected, and all other ones go to zero, the Rydberg medium behaves as deterministic 
near-perfect ``quantum scissors" \cite{qs}, cutting the initial coherent state into 
the subspace of Fock states with zero and one photon :
$$ | \alpha \rangle \rightarrow   | 0 \rangle + \alpha |1 \rangle $$
This is in agreement with the recent result of Kuzmich et al. \cite{kuzmich}. In addition, we have the 
explicit expressions to calculate the evolution between the initial Gaussian coherent state, and the final 
truncated non-Gaussian state, as illustrated on Fig.  \ref{wigner}.

As an exemple of realistic experimental parameters, let us consider  2500 atoms in
a sphere of  radius $R$ = 10  $\mu$m   
so $n_{at} =  6 \; 10^{11}$ cm$^{-3}$. 
For the (repulsive) 70s state of Rb$^{87}$,  one has  $|C_6|/(2 \pi)$ = 880  GHz~$ \mu m^6$ \cite{RR},
so the time scaling is  $T_R = R^6/|C_6|$ = 180~ns.
An initial coherent state with amplitude $|\alpha| \sim 1$,  excited by a pulse of duration 
 0.72 ns ($0.004 \, T_R$) will evolve into
 a non-Gaussian state within a time of 720 ns ($4 \, T_R$), as shown on Fig.  \ref{wigner}. 
If the atoms are inside a low-finesse optical cavity,   these numbers also warrant a large cooperativity parameter C 
with high outcoupling  mirror transmission
$T_c$ ($C \sim 180$ for  $T_c  = 0.01$), which should in turn warrant a good recovery of the photonic state \cite{Gorshkov07,Vuletic,us}. 

As a conclusion, we have studied the evolution of  a coherent  (Dicke) Rydberg polariton state, 
under the effect of  Rydberg-Rydberg  collisions.  The non-linearities are clearly large enough to 
have an effect at the few-photon level, even outside the dipole blockade range. They are able  to turn 
an input coherent state into a non-Gaussian state, at the expense of significant losses due to the 
decoherence of  states containing more that one polariton. Whether or not such decoherence effects 
can be avoided in order to reach a high input-output recovery of photonic states is still an open question.
 Let us emphasize however that  this decoherence does not prevent the polariton state, once created,  
 to be remapped with high efficiency on a photonic state, and then analyzed using an homodyne detection \cite{us}. 
 This  may also provide an  interesting way to investigate Rydberg-Rydberg  collisions. 

%%%%%%
\section*{Appendix} 
%%%%%%%
{\bf (A1)} The Dicke state $|n \rangle$ corresponding to $n$ excited atoms among $N$ atoms is defined 
as the eigenstate with eigenvalues ($n - N/2$, $r(r+1)$) of the operators ($\hat J_{\vec{k}}^{(z)}$,  
$\hat J_{\overrightarrow  k}^2$), with $\hat J_{\vec{k}}^{(z)}  = \sum_{i=1}^N ( |e, i \rangle \langle e, i| -  |g, i \rangle \langle g, i| )/2 $, 
$\hat J_{\vec{k}}^{(+)} = \sum_{i=1}^N |e, i \rangle \langle g, i|  \exp ( i \vec{k}. \vec{R_i})$,  $\hat J_{\vec{k}}^{(-)} = (\hat J_{\vec{k}}^{(+)} )^\dagger$
and $\hat J_{\vec{k}}^2 =  (\hat J_{\vec{k}}^{(+)}  \hat J_{\vec{k}}^{(-)}  + \hat J_{\vec{k}}^{(-)} \hat J_{\vec{k}}^{(+)})/2 + (\hat J_{\vec{k}}^{(z)})^2 $, 
where $\vec{R_i}$ is the position of atom $i$, and $\vec{k}$ is the total  wave-vector of the exciting light. Symmetric Dicke  states 
are obtained for $r = N/2$, and non-symmetric ones for $r  \leq N/2-1$. 

\vskip 5mm

%%%%%%
%\section*{Appendix 2} 
%%%%%%%
{\bf (A2)} To estimate the action of the collisions 
during the short excitation phase of duration $\tau$, two points must be considered. 
The first one is that the excitation of a Rydberg atom creates a ``blockade sphere" around it, where 
a second atom cannot be excited. 
The second one is that the quantities $U_n$ calculated above should remain close to one during $\tau$, 
so that the excitation and the collisions act on different time scales. 
To evaluate the first correction  we can use  
the following correlation function $G_{ij}$ derived in \cite{jovica2010} for
two level atoms and small $\omega$
\begin{equation}\label{corr-function}
G_{ij}\equiv \frac{\langle\hat\sigma^{i}_{ee}\hat\sigma^{j}_{ee}\rangle}
{\langle\hat\sigma^{i}_{ee}\rangle\langle \hat\sigma^{j}_{ee}\rangle}
=\frac{4\left|\int_{t_{0}}^{\tau}dt_1\,e^{i
\kappa_{ij}t_1} \Omega(t_1)\omega(t_1)\right|^2}{|\omega(\tau)|^4}.
\end{equation}
This function prevents two excited atoms to be closer to each other than 
a distance $r_b$  with magnitude given by $C_6 \tau /r_b^6 \sim 1$, where 
 $\tau$ is the duration of the excitation pulse. It is thus clear that $r_b$ will be small if $\tau$ is small. 
 More quantitatively, the modification of Eq. (\ref{U2-scaling})  
would be to substitute $I_2(T)$ by $c(\tau,T)I_2(T)$, where  $c(\tau,T)$ is a correction factor
evaluated numerically.  For the parameters quoted in the text, 
we numerically get $c(\tau,\tau)\!\approx \!0.984\!$ at the end of the excitation pulse, 
and this effect is thus negligible.  Similarly, using the results in Appendix 3 one gets for $t = \tau/T_R$ = 0.004  the values 
$U_2 \approx 0.94 + 0.05 \, i $, $U_3 \approx  0.82 + 0.13 \, i$, $U_4 \approx 0.66 + 0.21\, i$, that is acceptably close to one 
if  $|\alpha|$ is small enough (see also Fig.  \ref{nn}).

\vskip 5mm

%%%%%%%%
%\section*{Appendix 3}
%%%%%%%%%%
{\bf (A3)} The analytical expression of $I_2(t)$ used in the curves on the previous page is : 
\begin{eqnarray*}
I_2(t)&=& -1 - 8  \;   \text{Exp}\left[\frac{i t}{64}\right]  + \frac{1}{32} \left(-\pi  t -i t \; \text{Ei}\left[\frac{i t}{64}\right] \right)  \nonumber \\
&+& \frac{(1+ i)}{2}  \sqrt{2 \pi t}   \left(i-\text{erfi}\left[\left(\frac{1}{8}+\frac{i}{8}\right) \sqrt{\frac{t}{2}}\right]\right) \nonumber \\
&+& \frac{3  t^{2/3} }{16} \left(1-i \sqrt{3}\right) \left(\Gamma \left[-\frac{2}{3},-\frac{i t}{64}\right]+3 \; \Gamma   \left[\frac{1}{3},-\frac{i t}{64}\right]\right) \nonumber \\
&\approx& (i-1) \sqrt{\frac{\pi t }{2}} + \frac{9}{32} (1 - i \sqrt{3}) \;  \Gamma \left[\frac{1}{3}\right] \;  t^{2/3} \nonumber
\end{eqnarray*}
where the second expression is a short-time approximation valid up to $t \sim 0.1$. 
Here $t$ is a scaled time related to the physical  time $T$ by $t = T/T_R$, where  
$T_R= R^6/|C_6|$, and $R$ is the radius of the spherical volume of the sample. 

\vskip 5mm

{\bf Acknowledgments} This work is supported by the ERC  Grant 246669 ``DELPHI'', 
by the European  ITN project ``COHERENCE", and by the RTRA project ``COCORYCO". 
We thank Robin C\^ot\'e, Rosa Tualle-Brouri and Andrew Hilliard for useful discussions. 

%%%%%%%%%%%%%%%%%%%%%%%%%%%%%%%%%%%%%%

%%%%%%%%%%%%%%%%%%%%%%%%%%%%%%%%%%%%%%%%

\end{document}